\documentclass{sig-alternate}

\usepackage[usenames,dvipsnames]{color}
\usepackage{enumerate}
\usepackage{algorithm}
\usepackage{subfigure}
\usepackage{amssymb}
\usepackage{graphicx}
\usepackage{hyperref}
\usepackage{xspace}

\newdef{definition}{Definition}

\DeclareGraphicsExtensions{.eps,.pdf,.jpg,.png}

\definecolor{black}{RGB}{0, 0, 0}
\definecolor{gray}{RGB}{121, 121, 121}
\definecolor{blue}{RGB}{0, 84, 147}
\definecolor{green}{RGB}{146, 144, 0}
\definecolor{mocha}{RGB}{147, 82, 0}
\definecolor{asparagus}{RGB}{146, 144, 0}

\newcommand{\links}[0]{L}
\newcommand{\nodes}[0]{N}
\newcommand{\graph}[0]{G}
\newcommand{\neighbors}[0]{\Gamma}

\newcommand{\n}[0]{n}
\newcommand{\m}[0]{m}
\newcommand{\degree}[0]{k}
\newcommand{\ndegree}[0]{\degree_N}
\newcommand{\mixing}[0]{r}
\newcommand{\distance}[0]{l}
\newcommand{\clustering}[0]{C}
\newcommand{\modularity}[0]{Q}

\newcommand{\burning}[0]{p}
\newcommand{\burned}[0]{s}
\newcommand{\linking}[0]{q}

\newcommand{\FF}[0]{Forest Fire\xspace}
\newcommand{\BTF}[0]{Butterfly\xspace}
\newcommand{\CPY}[0]{Copying\xspace}
\newcommand{\CIT}[0]{Citation\xspace}

\newcommand{\cora}[0]{{\itshape Cora}\xspace}

\newcommand{\secref}[1]{Section~\ref{sec:#1}\xspace}

\newcommand{\figref}[1]{Figure~\ref{fig:#1}\xspace}
\newcommand{\figsref}[2]{Figures~\ref{fig:#1} and~\ref{fig:#2}\xspace}
\newcommand{\tblref}[1]{Table~\ref{tbl:#1}\xspace}

\newcommand{\equref}[1]{equation~(\ref{equ:#1})\xspace}
\newcommand{\equsref}[2]{equations~(\ref{equ:#1}) and~(\ref{equ:#2})\xspace}

\newcommand{\HLINE}[0]{\noalign{\hrule height 1pt}}
\newcommand{\plotwidth}[0]{0.38\columnwidth}
\newcommand{\drawwidth}[0]{0.42\columnwidth}
\newcommand{\figurewidth}[0]{0.52\columnwidth}

\begin{document}


\title{Model of Complex Networks based on Citation Dynamics}

\numberofauthors{2}
\author{
\alignauthor
	Lovro \v{S}ubelj\\
	\affaddr{University of Ljubljana}\\
	\affaddr{Faculty of Computer and Information Science}\\
	\affaddr{Tr\v{z}a\v{s}ka cesta 25, SI-1000 Ljubljana, Slovenia}\\
	\email{lovro.subelj@fri.uni-lj.si}
\alignauthor
	Marko Bajec\\
	\affaddr{University of Ljubljana}\\
	\affaddr{Faculty of Computer and Information Science}\\
	\affaddr{Tr\v{z}a\v{s}ka cesta 25, SI-1000 Ljubljana, Slovenia}\\
	\email{marko.bajec@fri.uni-lj.si}
}

\conferenceinfo{LSNA'13,}{May 14, 2013, Rio de Janeiro, Brazil.}
\crdata{X-XXXXX-XX-X/XX/XX} 
\CopyrightYear{2013}

\date{23 March 2013}

\maketitle


\begin{abstract}
Complex networks of real-world systems are believed to be controlled by common phenomena, producing structures far from regular or random. These include scale-free degree distributions, small-world structure and assortative mixing by degree, which are also the properties captured by different random graph models proposed in the literature. However, many (non-social) real-world networks are in fact disassortative by degree. Thus, we here propose a simple evolving model that generates networks with most common properties of real-world networks including degree disassortativity. Furthermore, the model has a natural interpretation for citation networks with different practical applications.
\end{abstract}





\section{Introduction} \label{sec:intro}
Networks are the simplest representation of complex systems of interacting parts. Examples of these are ubiquitous in practice, including large social networks~\cite{FIA11}, information systems~\cite{SB11s} and cooperate ownerships~\cite{VGB11}, to name just a few. Despite a seemingly plain form, real-world networks reveal characteristic structural properties that are absent from regular or random systems~\cite{WS98,BA99}. Thus, networked systems are believed to be controlled by common phenomena.

Scale-free degree distributions~\cite{BA99}, small-world phenomena~\cite{WS98}, degree mixing~\cite{New02} (i.e., degree correlations at links' ends) and existence of communities~\cite{GN02} (i.e., densely linked groups of nodes) are perhaps among most widely analyzed properties of large real-world networks. Note that community structure implies assortative (i.e, positively correlated) mixing by degree~\cite{NP03}, which can be seen as a tendency of hubs (i.e., highly linked nodes) to cluster together. The above are also the properties captured by many random graph models proposed in the literature~\cite{KR05,LKF07,MAF08,WH09}.

\begin{figure}[b]
\centering
\includegraphics[width=\figurewidth]{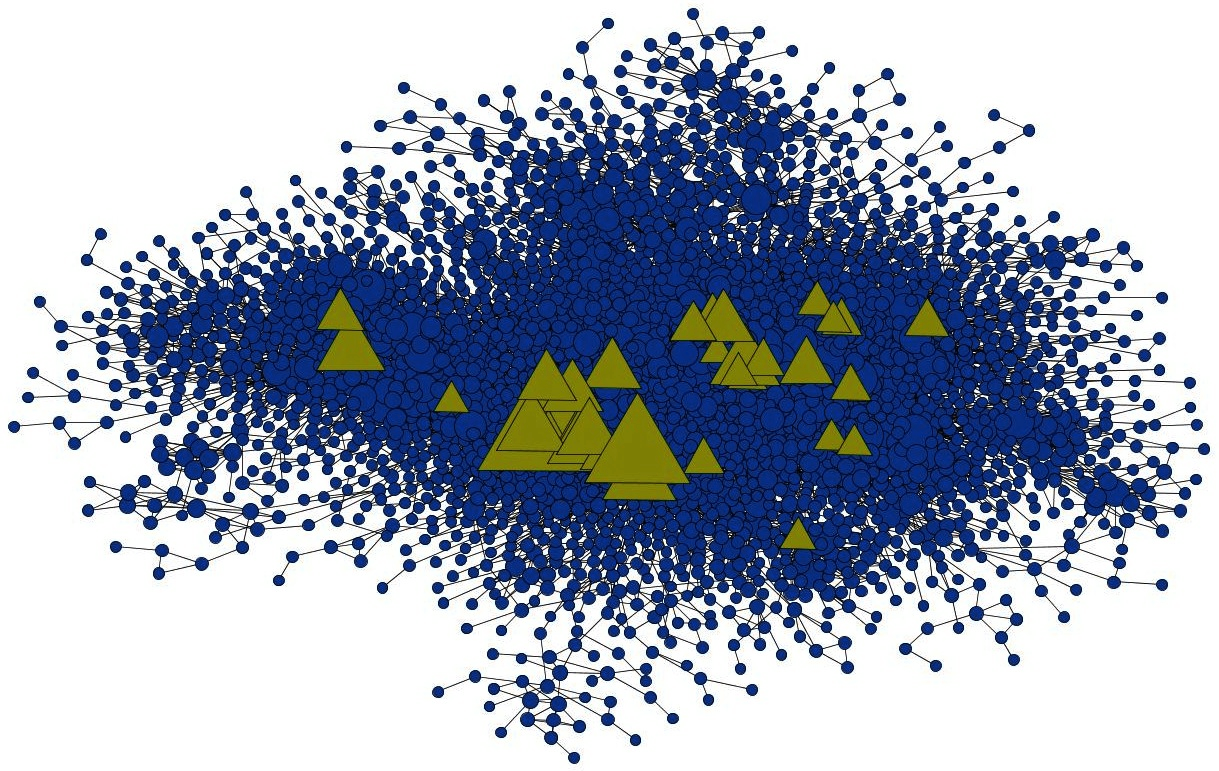}
\caption{\label{fig:cora}Data mining part of \cora citation network~\cite{MNRS00} with highlighted hubs (i.e., $1\%$ of most highly linked nodes) that are scattered across the network. (Node sizes are proportional to degrees.)}
\end{figure}

However, most (non-social) networks deviate from this figure. Biological and technological networks are in fact degree disassortative (i.e., negatively correlated), while different information networks often reveal no clear degree mixing~\cite{New02,HL11} (see~\figref{cora}). Thus, we here propose an evolving random graph model based on the link copying mechanism~\cite{KR05}. Each newly added node explores the network using the burning process in~\cite{LKF07}, while links of the visited nodes are copied independently of the latter. The model generates scale-free small-world networks with community structure and also degree disassortativity. Furthermore, it has a natural interpretation for citation networks. The above process imitates an author of a paper including references into the bibliography (i.e., its citation dynamics), which enables different practical applications in bibliometrics (see~\secref{cora}).

The rest of the paper is structured as follows. \secref{model} introduces the proposed (\CIT) model, while a thorough analysis is given in \secref{experiments}. \secref{conclusion} concludes the paper.


\begin{figure*}[t]
\centering
\subfigure[\label{fig:models:FF}\FF model]{
\includegraphics[width=\drawwidth]{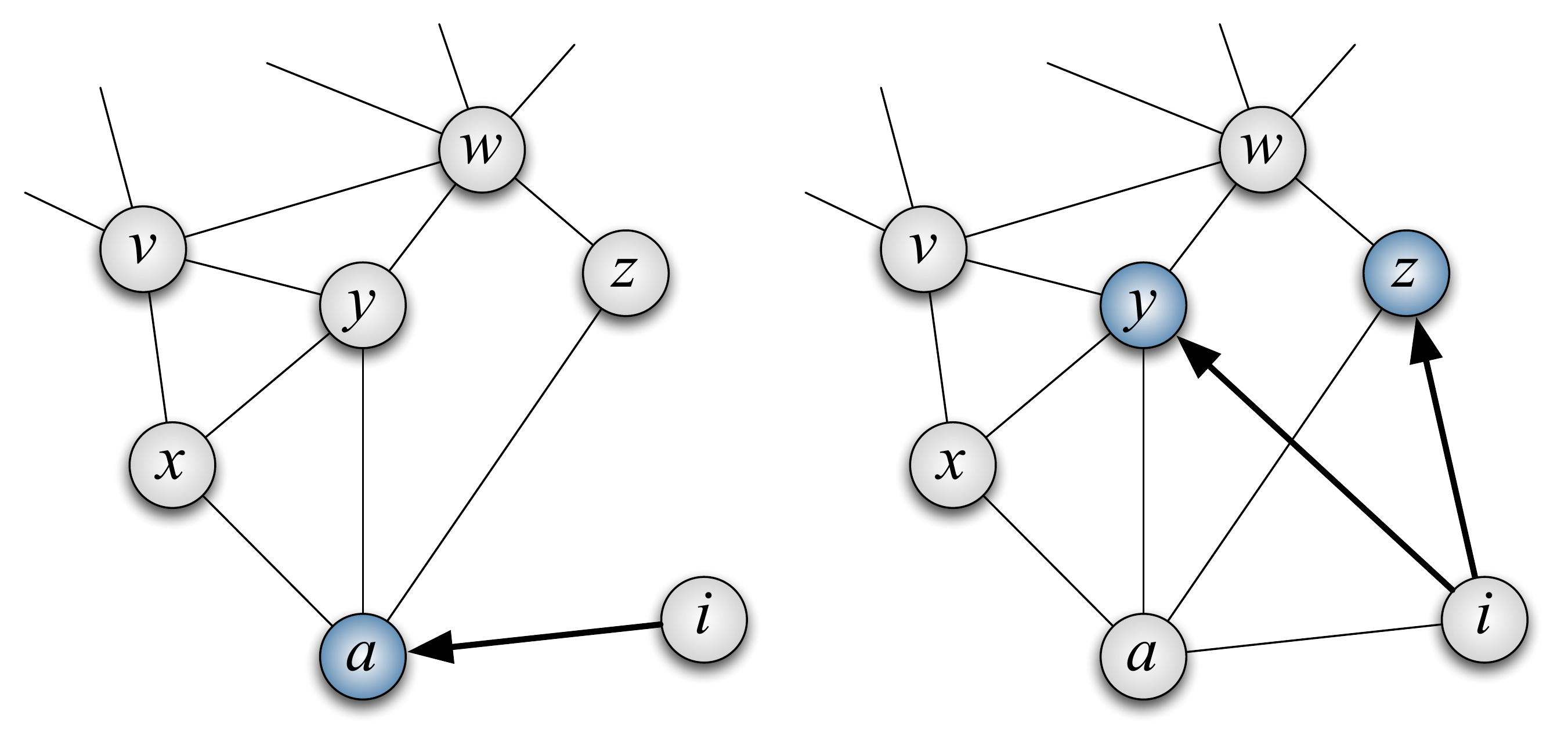}}
\subfigure[\label{fig:models:BTF}\BTF model]{
\includegraphics[width=\drawwidth]{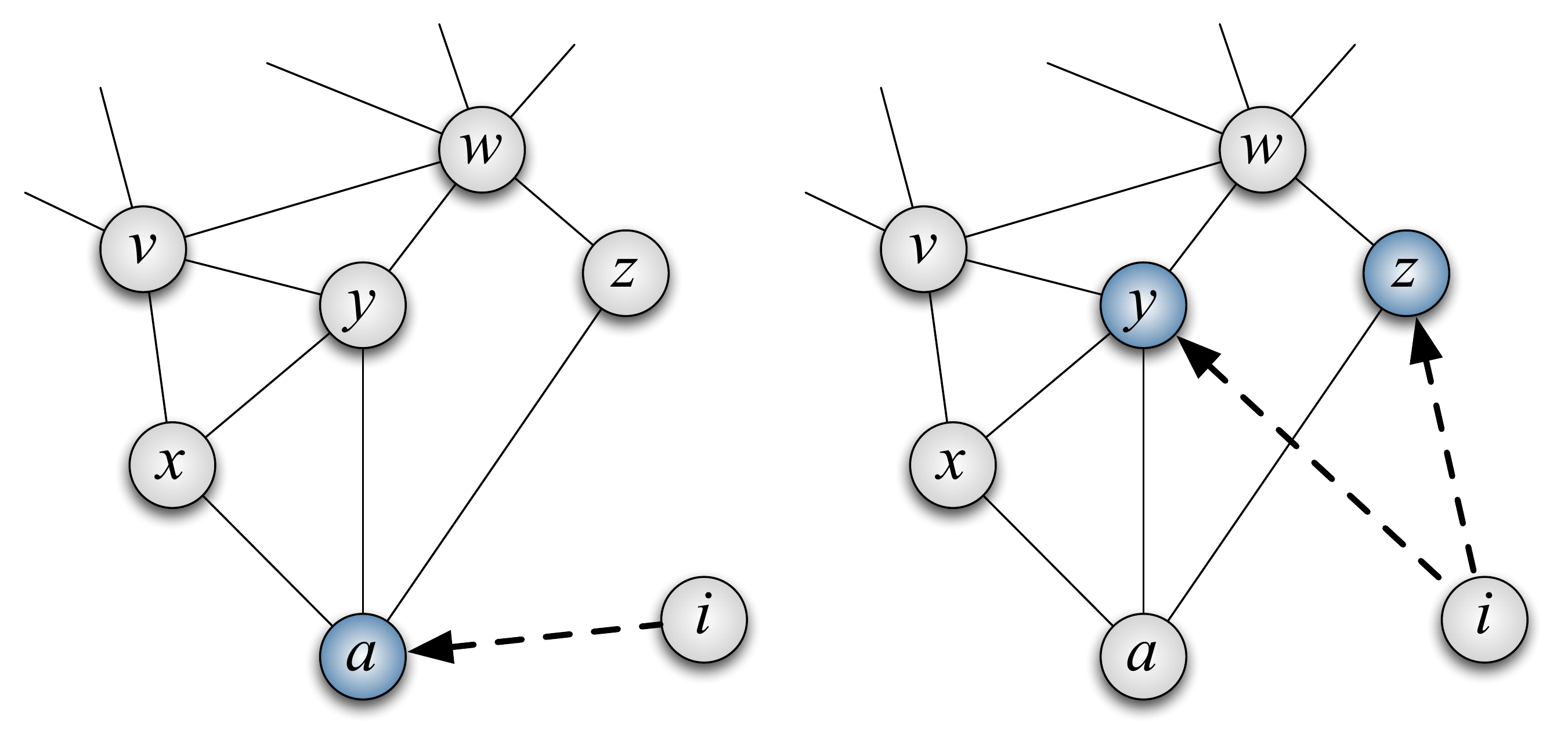}}
\subfigure[\label{fig:models:CPY}\CPY model]{
\includegraphics[width=\drawwidth]{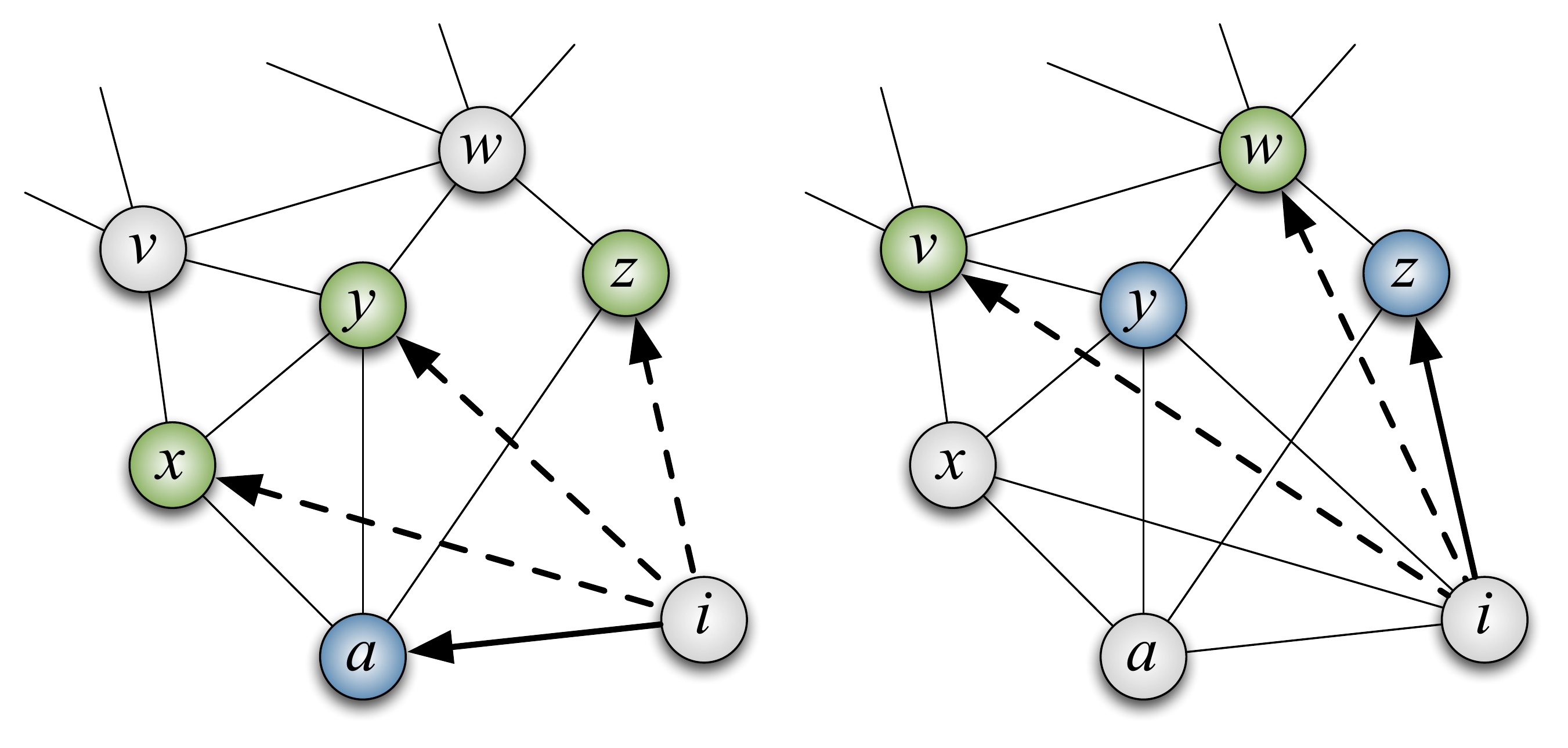}}
\subfigure[\label{fig:models:CIT}\CIT model (our)]{
\includegraphics[width=\drawwidth]{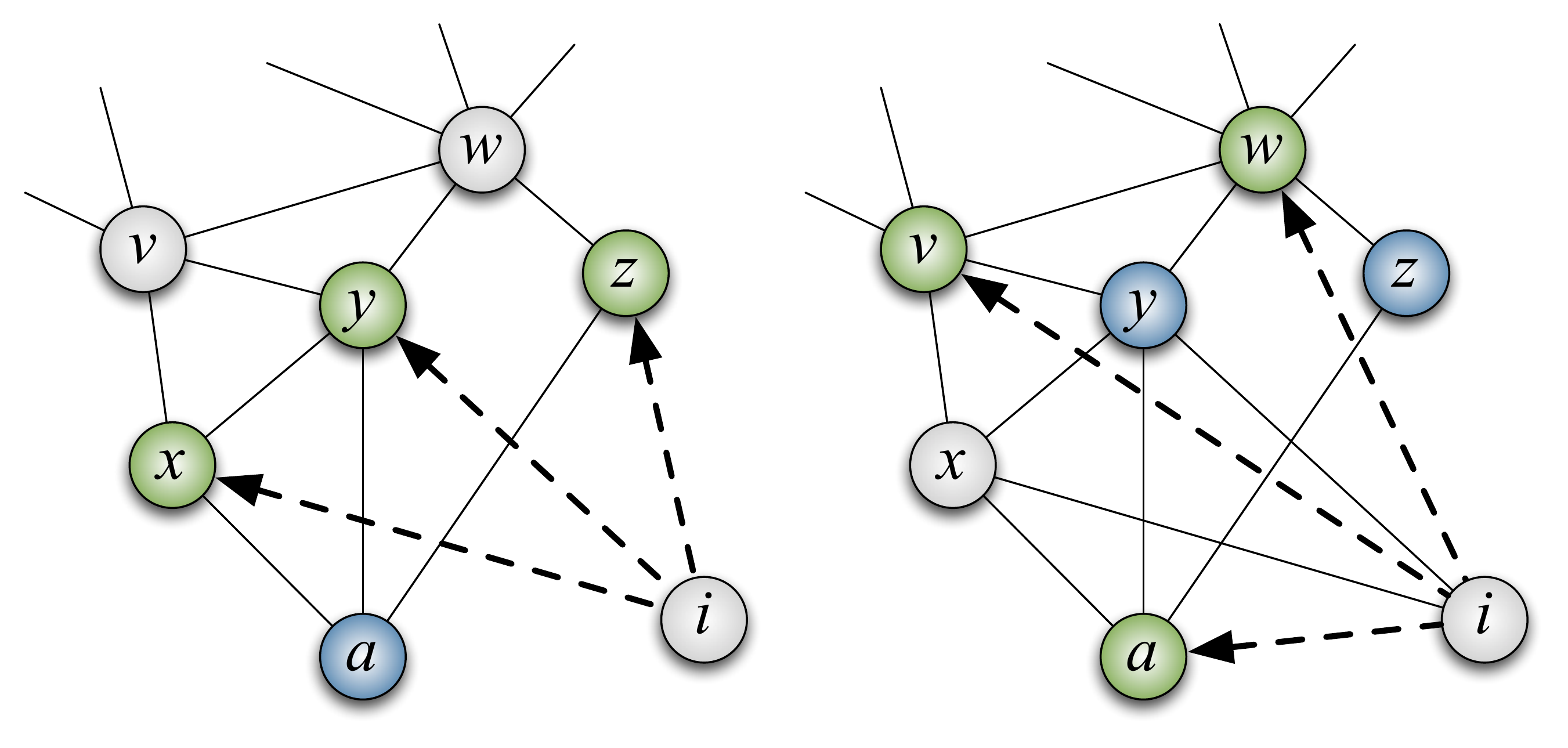}}
\caption{\label{fig:models}Schematic representation of linking dynamics of different graph models.
(a)~In \FF model~\cite{LKF07}, newly added node $i$ selects an ambassador $a$ (blue node) uniformly at random and links to it (solid arrow). Next, some of its neighbors are taken as the ambassadors (e.g., $y$ and $z$) and the process repeats. (b)~\BTF model~\cite{MAF08} forms links only with some fixed probability (dashed arrows). (c)~In \CPY model~\cite{KR05}, node $i$ links to $a$ and also to some of its neighbors $x$, $y$, $z$ (green nodes). (d)~Proposed \CIT model forms links only with the neighbors of the ambassador $a$ (e.g., $x$ and $y$), however, $i$ can still link to $a$.}
\end{figure*}

\section{The citation model} \label{sec:model}
Let a network be represented by a simple graph $\graph(\nodes,\links)$, where $\nodes$ is the set of nodes, $|\nodes|=\n$, and $\links$ is the set of links, $|\links|=\m$. Next, let $\neighbors_i$ be the set of neighbors of node $i\in\nodes$ and let $\degree_i$ be its degree, $\degree_i=|\neighbors_i|$. Last, let $\degree$ be the mean degree and $\ndegree$ the mean neighbor degree.

Proposed graph model is based on the burning process of \FF model~\cite{LKF07}, which we introduce first. Due to simplicity, the model is presented for undirected networks.

Let $\burning$ be the burning probability, $\burning\in [0,\frac{1}{2})$ (see below). Initially, the network consists of a single node, while for each newly added node $i$, the burning process proceeds as follows.
\begin{enumerate}[(1)]
\item $i$ chooses an ambassador $a\in\nodes$ uniformly at random (we say that $i$ burns $a$) and links to it.
\item $i$ randomly selects (at most) $x_\burning$ neighbors of $a$ that were not yet burned $a_1,\dots,a_{x_\burning}\in\neighbors_a$ and links to them. ($x_\burning$ is sampled from a geometric distribution with mean $\frac{\burning}{1-\burning}$.)
\item $a_1,\dots,a_{x_\burning}$ are taken as the ambassadors of $i$ (step~(2)).
\end{enumerate}
Since each node can be visited at most once, the burning process surely converges. Thus, to generate a network with $\n$ nodes, the model repeats the above procedure $\n-1$ times. 

\FF model produces shrinking diameters and densification phenomena observed in temporal networks~\cite{LKF07}. Furthermore, generated networks are scale-free and small-world, and reveal a pronounced community structure. However, in contrast to many real-world networks, the model gives degree assortative networks (see~\secref{experiments}).

The model also has a natural interpretation for citation networks. Burning process imitates an author of a paper including references into the bibliography (i.e., citation dynamics). Author first reads a related paper, or selects the paper that triggered the research, and cites it (step~(1)). Author then considers its bibliography for other related papers (step~(2)). Some of these are further considered and also cited, while the author continues as before (step~(3)). Nevertheless, \FF~model fails to reproduce some of the properties of citation networks (e.g., degree mixing).

Note that the described process assumes that authors read, or at least consider, all the papers they cite. However, this is indeed not the case~\cite{SR03}. For example, seminal work on random graphs conducted by Erd\"{o}s and R\'{e}nyi~\cite{ER59} is perhaps among most widely cited papers in network science literature. Although, presumably, only a smaller number of authors have actually read the original paper. As the work is widely discussed elsewhere, most authors have just copied the reference from another paper. On the other hand, authors also do not cite all the papers they read, though related to their work. This can be simply due to space limitations. Nevertheless, a paper can still be read thoroughly, with many of its references further considered and cited. 

Examples suggest that the papers that authors read or cite are selected due to two, not necessarily dependent, processes. We thus propose a \CIT model that adopts the above burning procedure to traverse the network, while the links are formed according to another independent process.

Let $q$ be the linking probability, $q\in [0,1)$ (see below). Initially, the network consists of a single link, while for each newly added node $i$, the model proceeds as follows.
\begin{enumerate}[(1)]
\item $i$ chooses an ambassador $a\in\nodes$ uniformly at random.
\item $i$ randomly selects (at most) $x_\burning$ neighbors of $a$ that were not yet burned $a_1,\dots,a_{x_\burning}\in\neighbors_a$.
\item $i$ randomly selects (at most) $x_\linking$ neighbors of $a$ that were not yet linked $j_1,\dots,j_{x_\linking}\in\neighbors_a$ and links to them. 
\item $a_1,\dots,a_{x_\burning}$ are taken as the ambassadors of $i$ (step~(2)).
\end{enumerate}
Details are the same as before. Again, the process surely converges, while the entire procedure is repeated $\n-2$~times.

Let $\burned$ be the mean number of burned nodes (i.e., ambassadors). A node selects $\frac{\burning}{1-\burning}$ ambassadors on each step, thus,
\begin{equation}
\burned \leq \sum_{x=0}^{\infty} \left(\frac{\burning}{1-\burning}\right)^x \leq \frac{1-\burning}{1-2\burning}.
\label{equ:burned}
\end{equation}

A node will fail to form any link (i.e., become isolated) with probability $(1-\linking)^\burned$. Although isolated nodes are a common property of real-world networks, they are often ignored in practice or the network is even reduced to the largest connected component. Thus, for the analysis here, we repeat the procedure until the largest component has $\n$ nodes.

Since a node forms $\frac{\linking}{1-\linking}$ links on each step, expected network degree is (with $1-\left(1-\linking\right)^{\burned}$ correction for isolated~nodes)
\begin{equation}
\degree \leq \frac{2\linking\burned}{1-\linking-\left(1-\linking\right)^{\burned+1}}.
\label{equ:degree}
\end{equation}

\begin{figure}[h]
\centering
\subfigure[\label{fig:analysis:p:v}Ambassadors $\burned$]{
\includegraphics[width=\plotwidth]{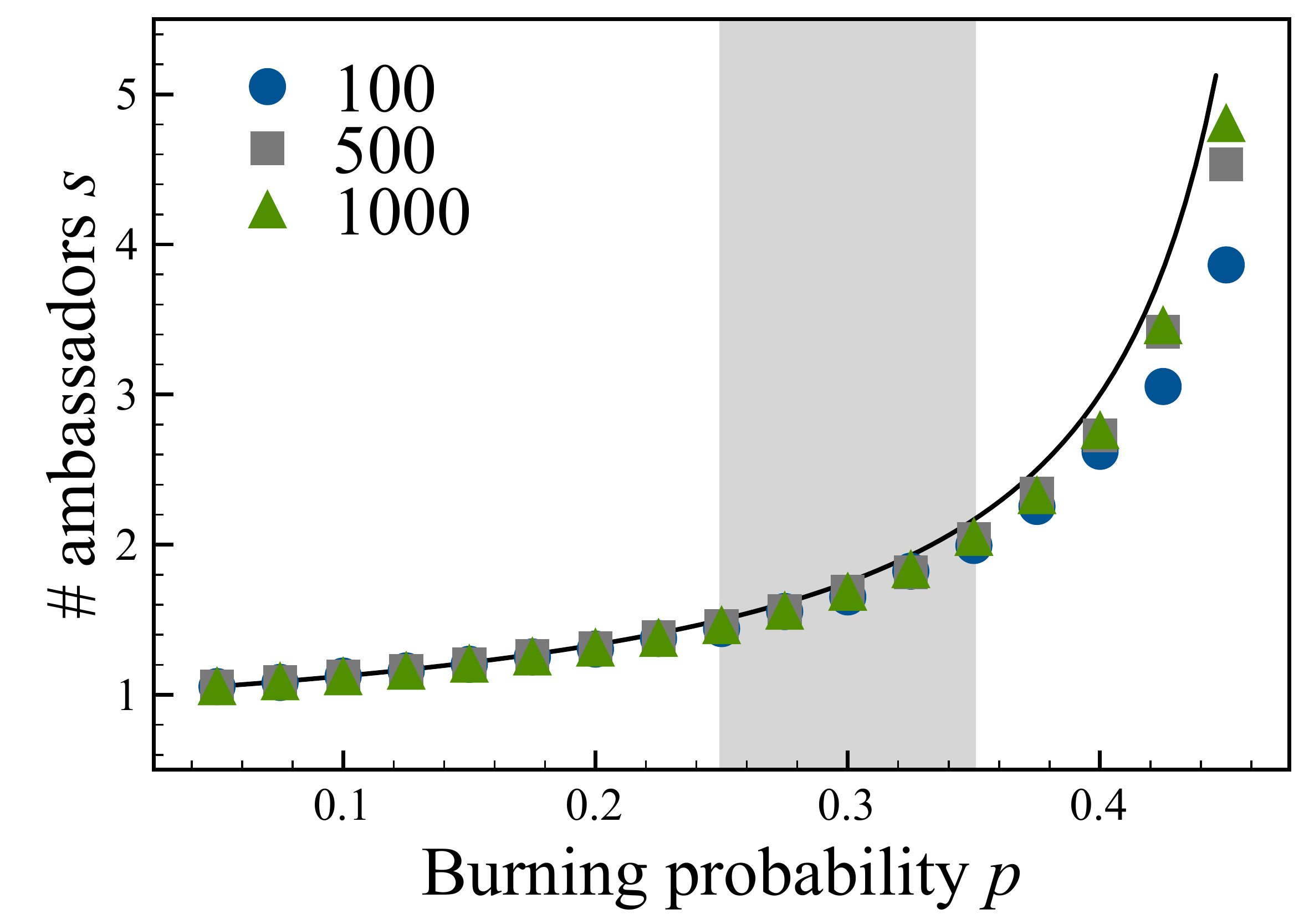}}
\subfigure[\label{fig:analysis:q:k}Network degree $\degree$]{
\includegraphics[width=\plotwidth]{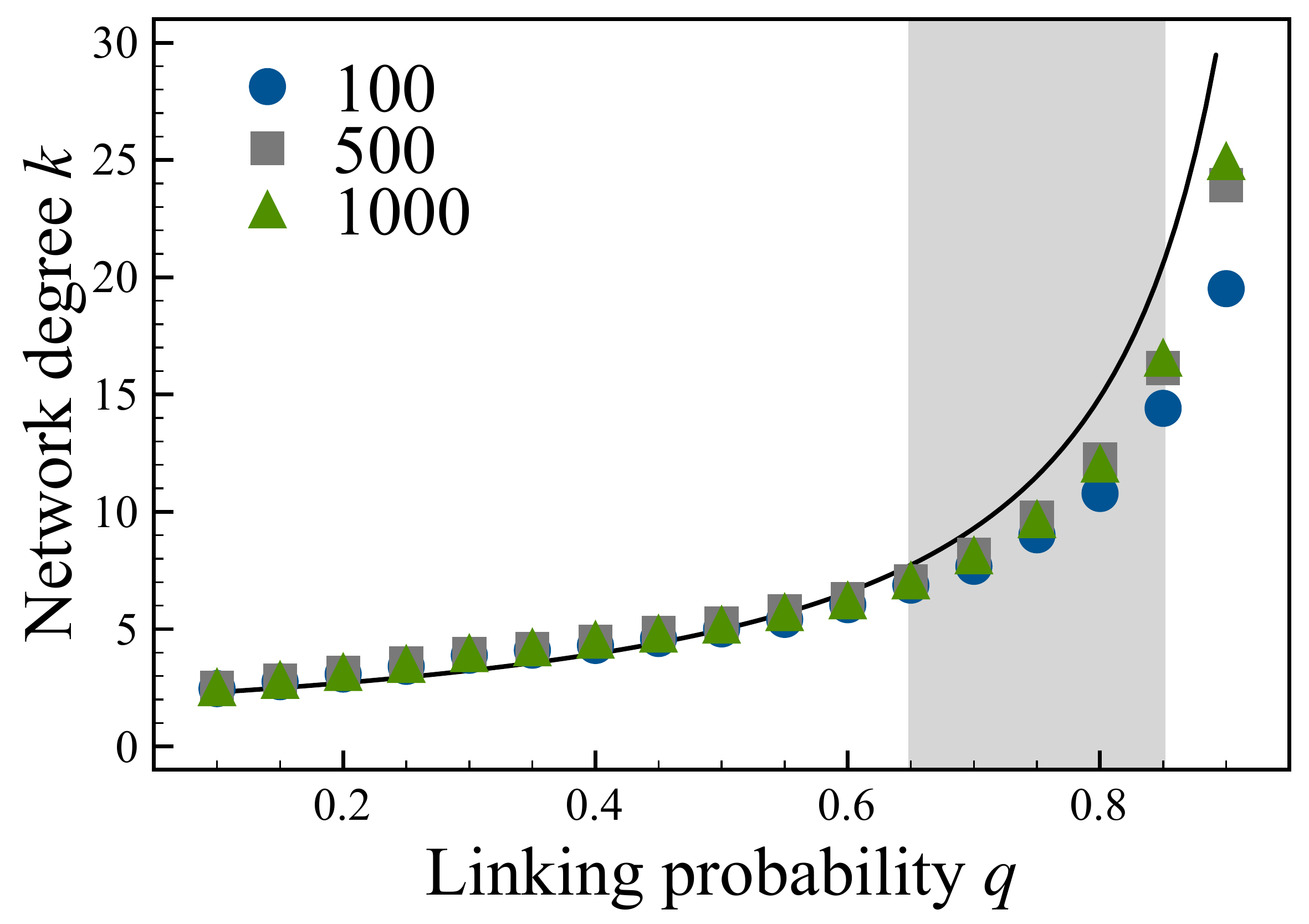}}
\caption{\label{fig:analysis}Analysis of \CIT model at different $\burning$ and $\linking=0.75$ (left), and $\burning=0.3$ and different $\linking$ (right). Solid lines show theoretical bounds in~\equsref{burned}{degree}. (Results are estimates of the mean over $100$ network realizations with different $\n$. Shaded regions correspond to likely parameter values~\cite{LJTBH11}.)}
\end{figure}

\begin{figure*}[t]
\centering
\subfigure[\label{fig:comparison:p:k}Network degree $\degree$]{
\includegraphics[width=\plotwidth]{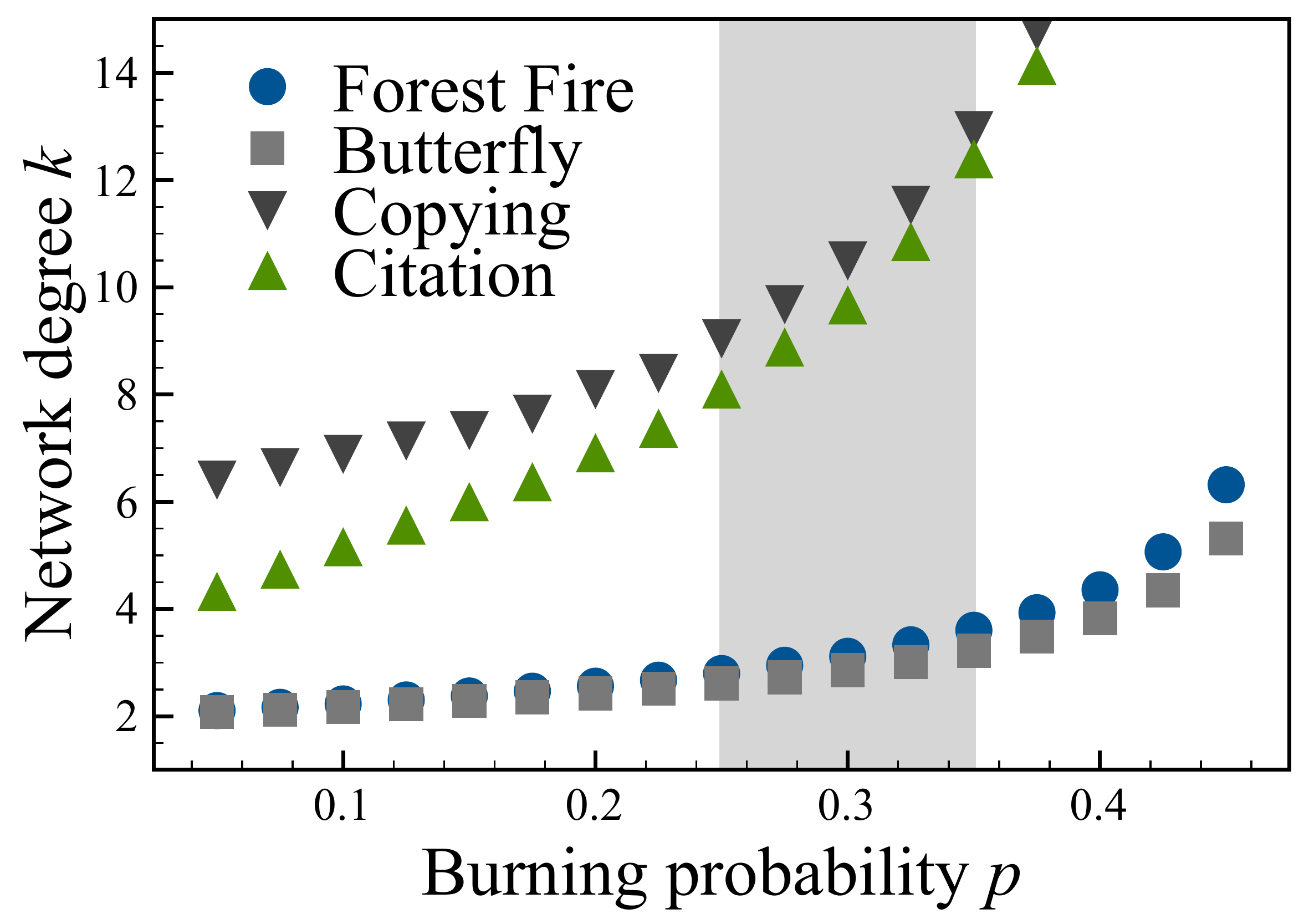}}
\subfigure[\label{fig:comparison:p:r}Degree mixing $\mixing$]{
\includegraphics[width=\plotwidth]{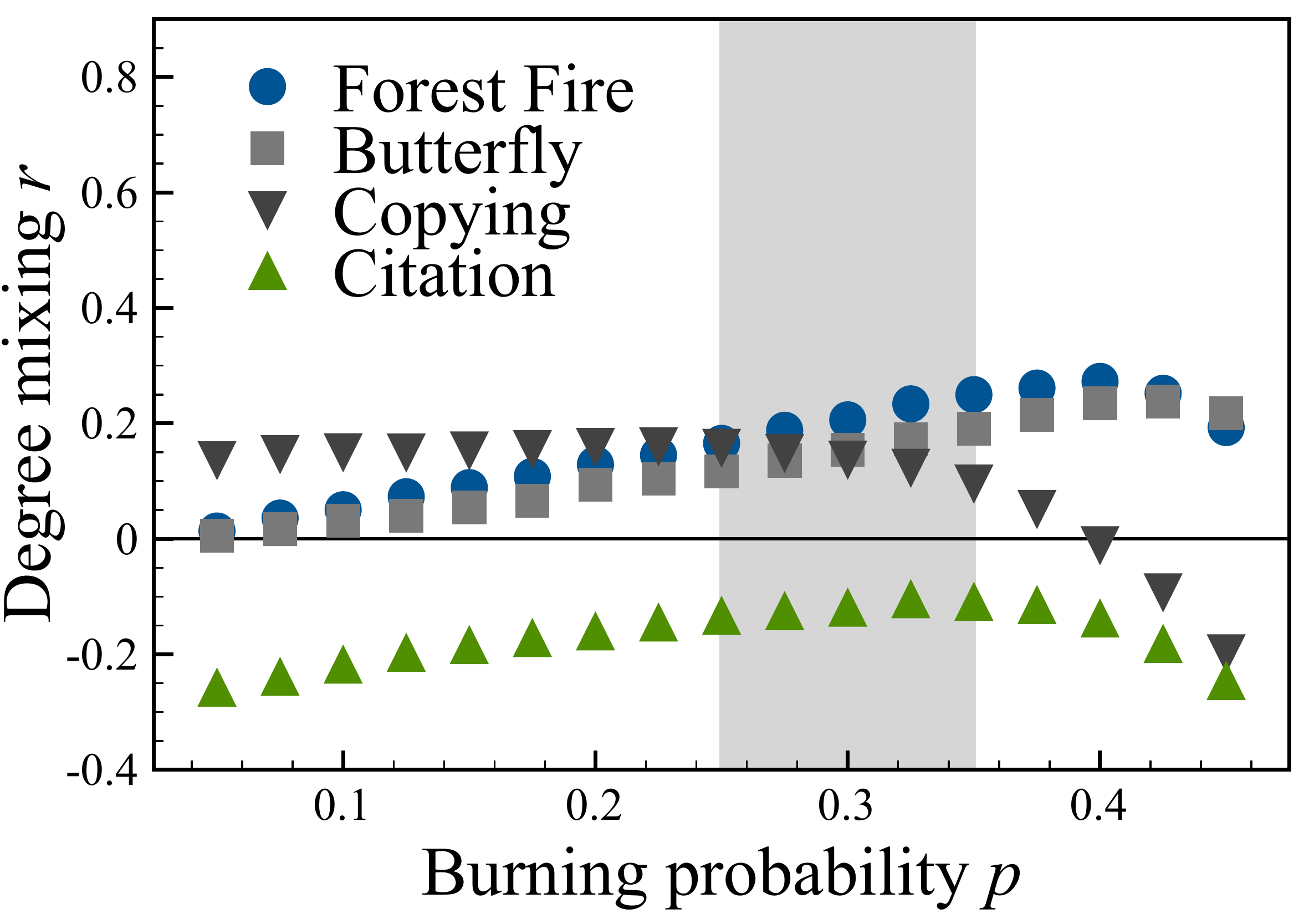}}
\subfigure[\label{fig:comparison:p:l}Mean distance $\distance$]{
\includegraphics[width=\plotwidth]{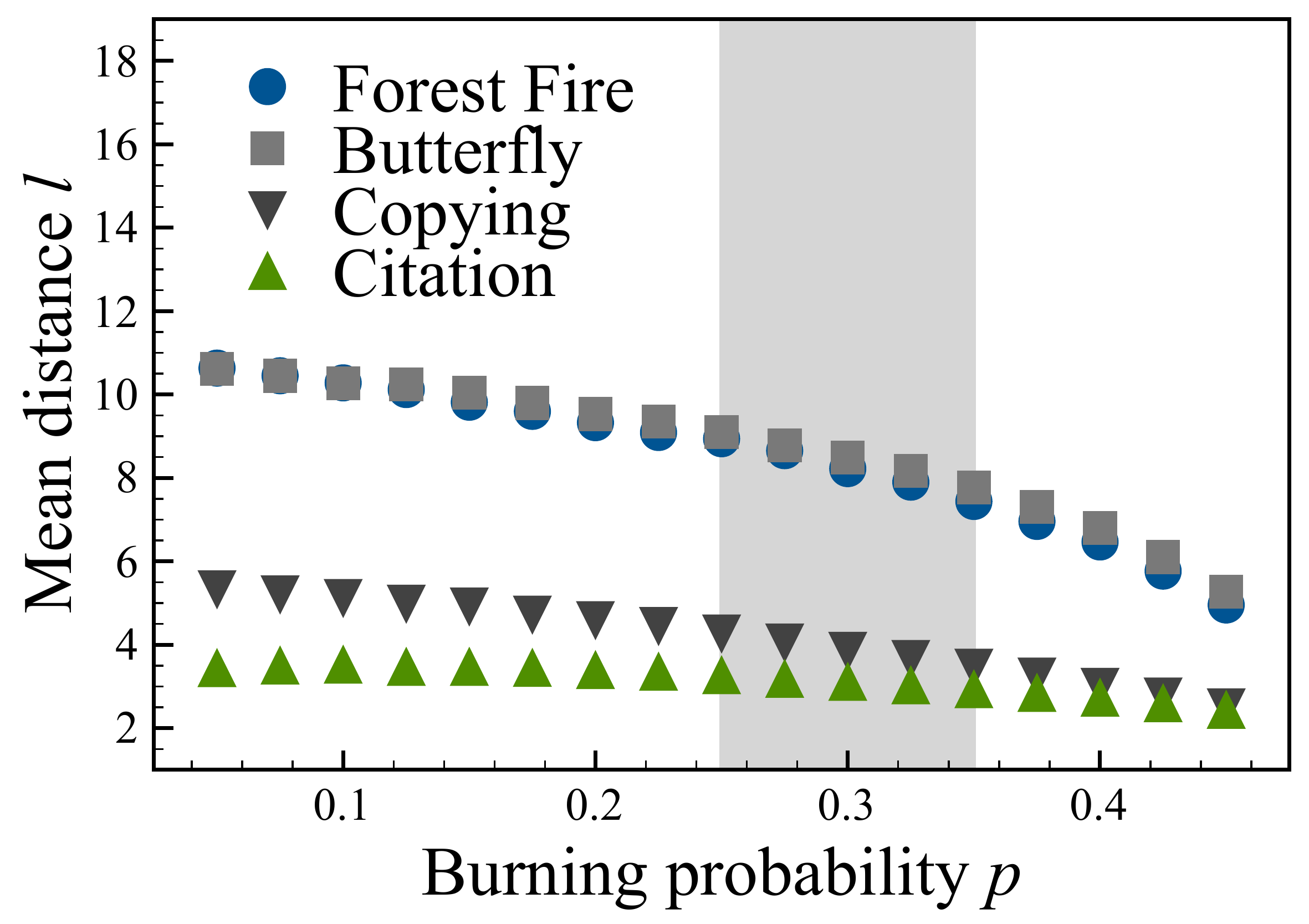}}
\subfigure[\label{fig:comparison:p:C}Clustering~$\clustering$]{
\includegraphics[width=\plotwidth]{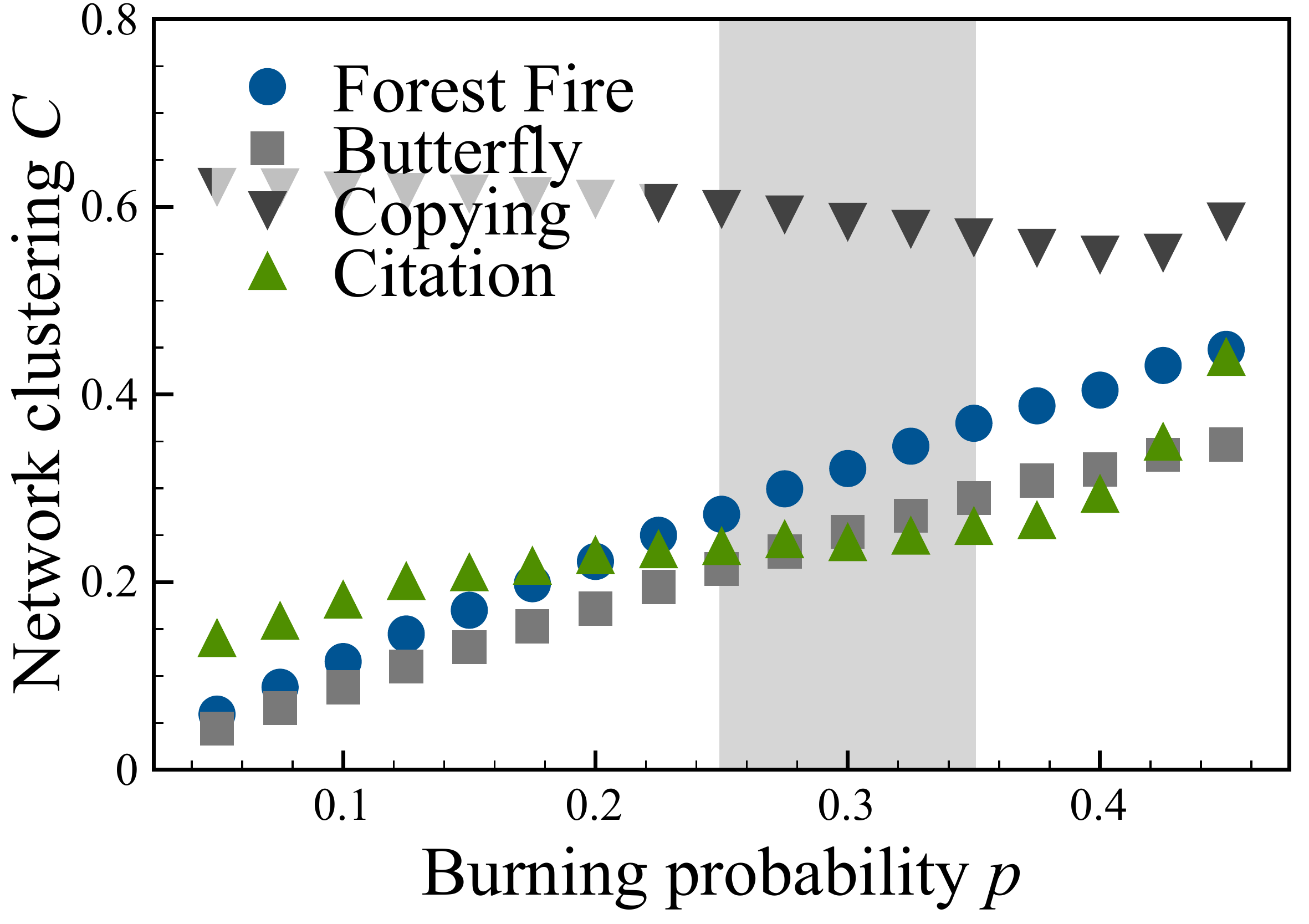}}
\subfigure[\label{fig:comparison:p:Q}Modularity $\modularity$]{
\includegraphics[width=\plotwidth]{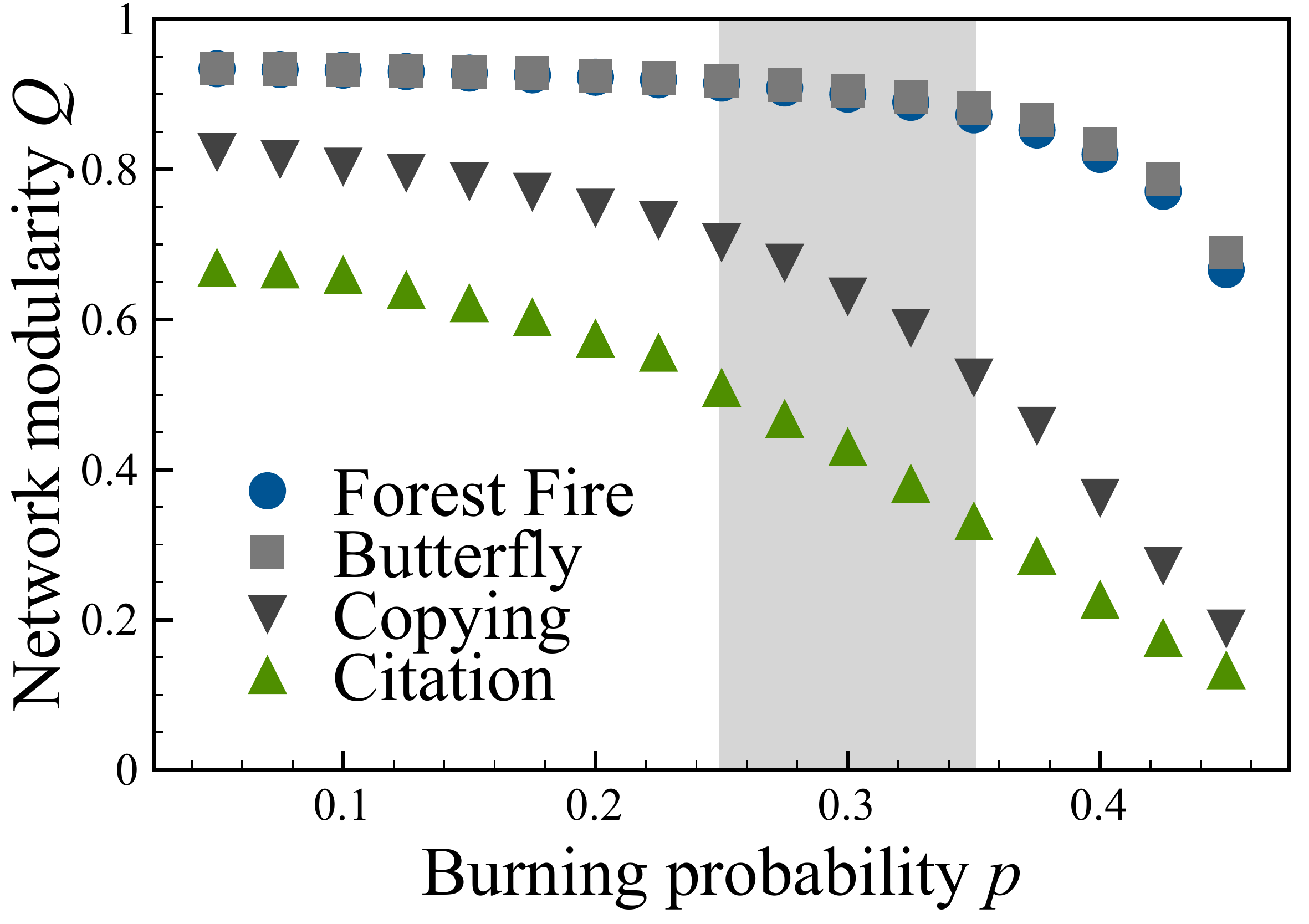}}
\subfigure[\label{fig:comparison:q:k}Network degree $\degree$]{
\includegraphics[width=\plotwidth]{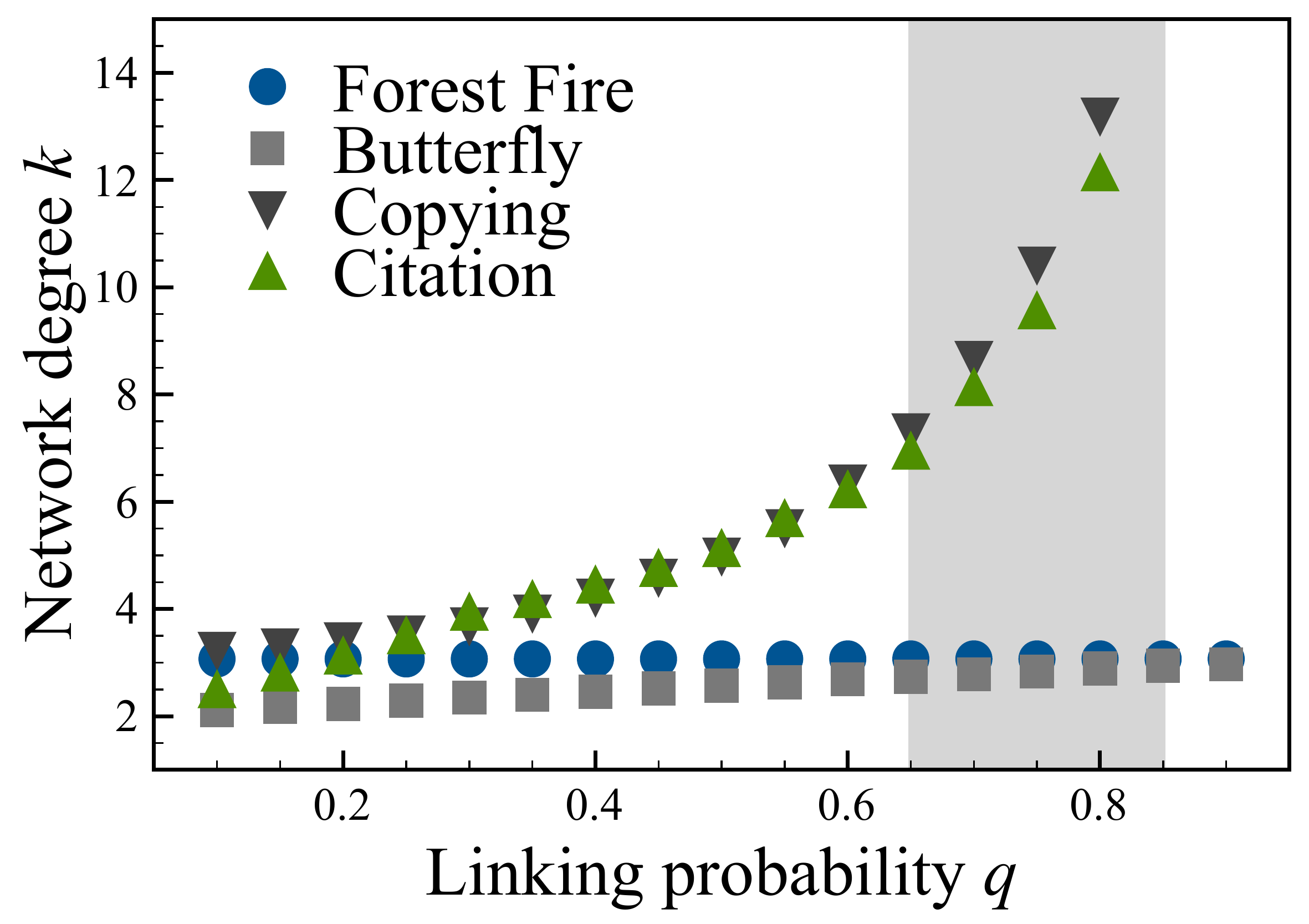}}
\subfigure[\label{fig:comparison:q:r}Degree mixing $\mixing$]{
\includegraphics[width=\plotwidth]{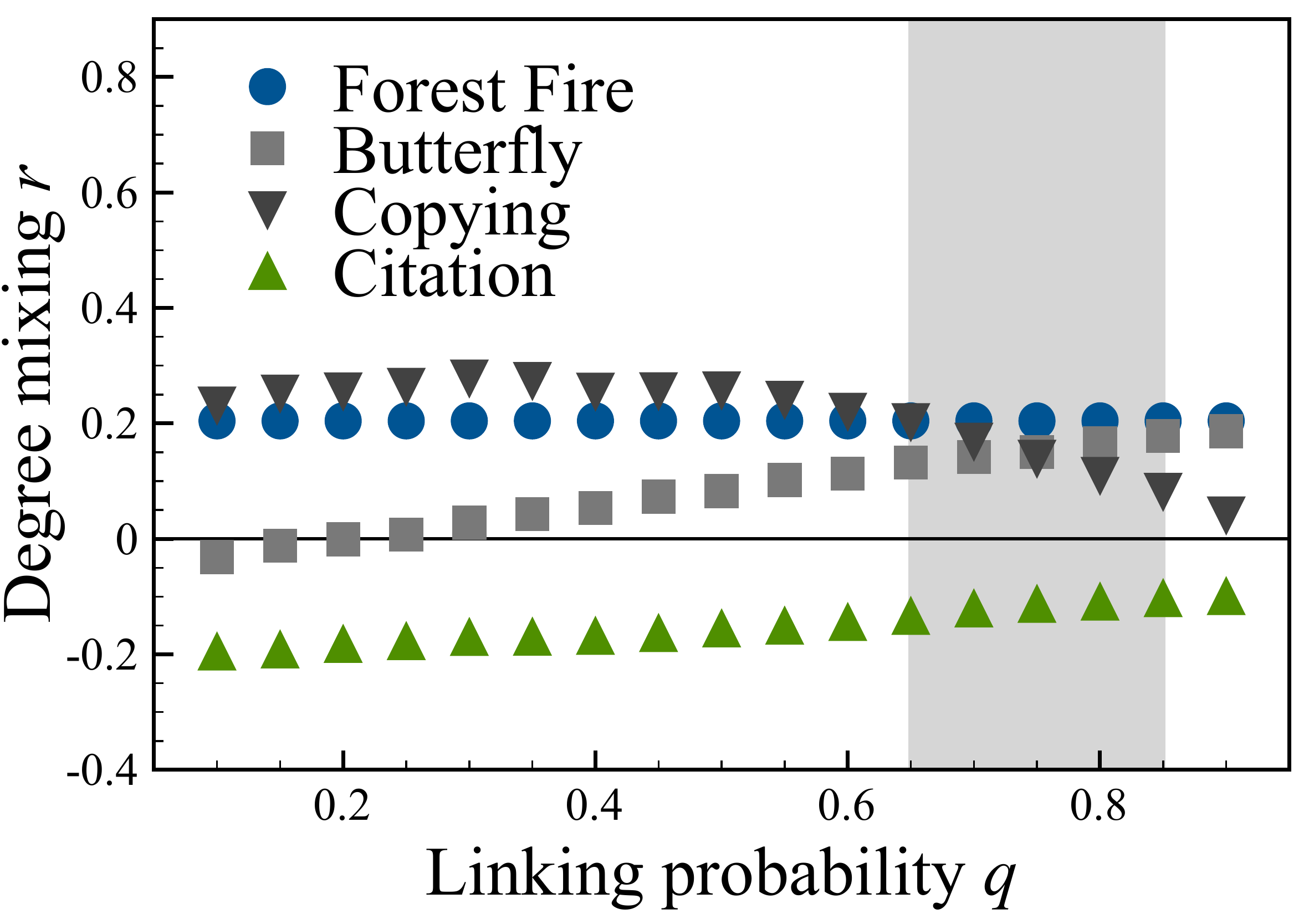}}
\subfigure[\label{fig:comparison:q:l}Mean distance $\distance$]{
\includegraphics[width=\plotwidth]{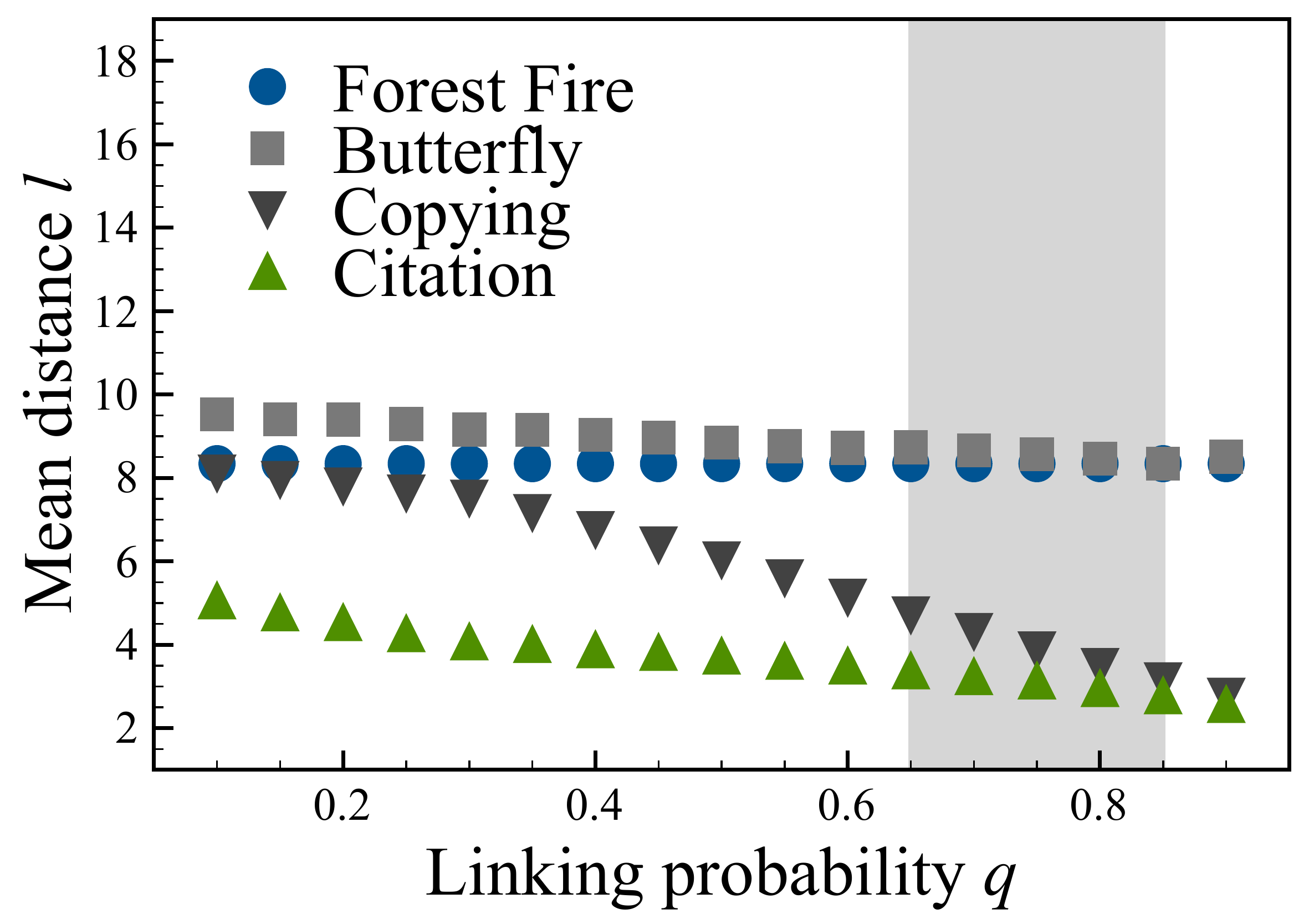}}
\subfigure[\label{fig:comparison:q:C}Clustering~$\clustering$]{
\includegraphics[width=\plotwidth]{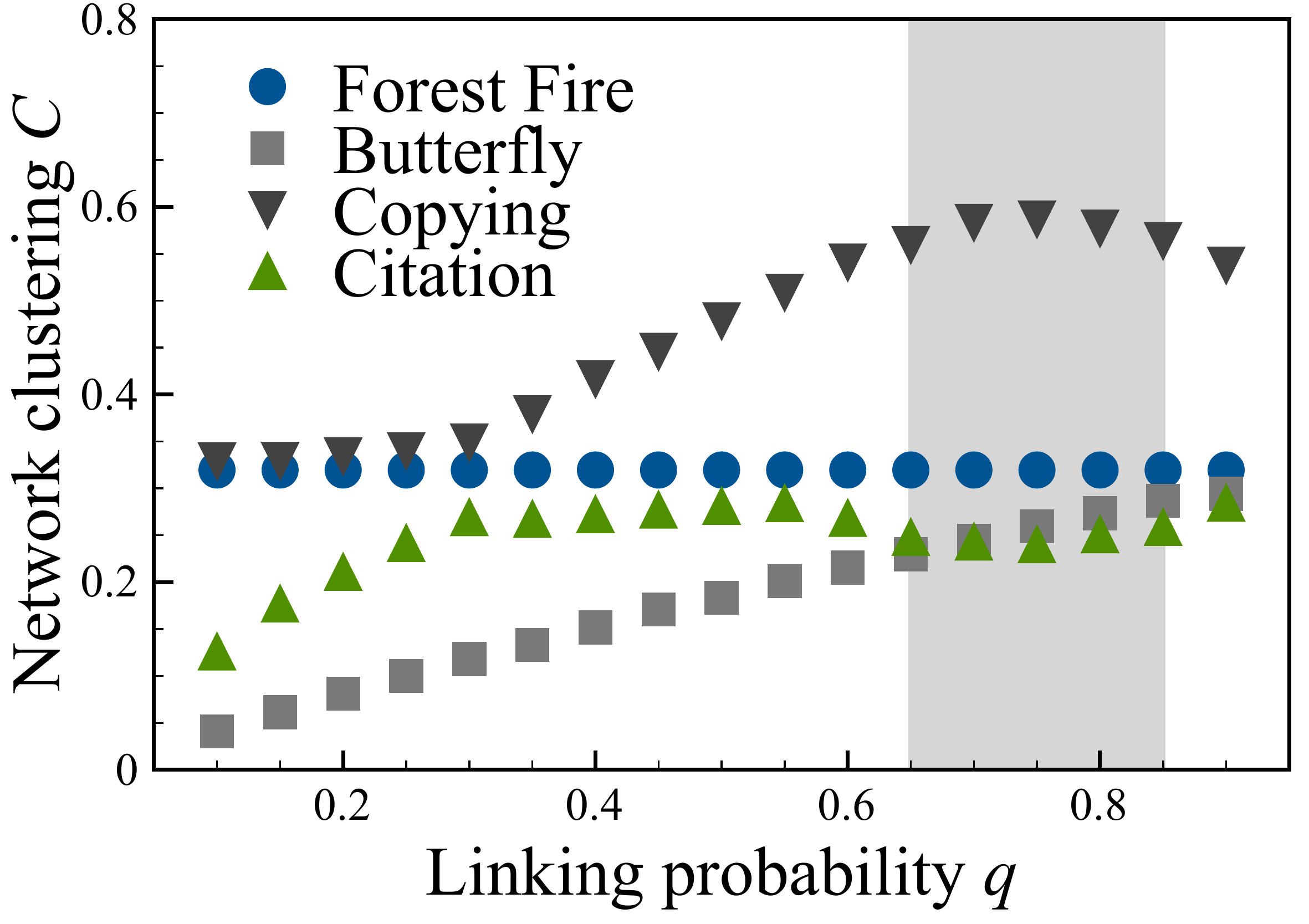}}
\subfigure[\label{fig:comparison:q:Q}Modularity $\modularity$]{
\includegraphics[width=\plotwidth]{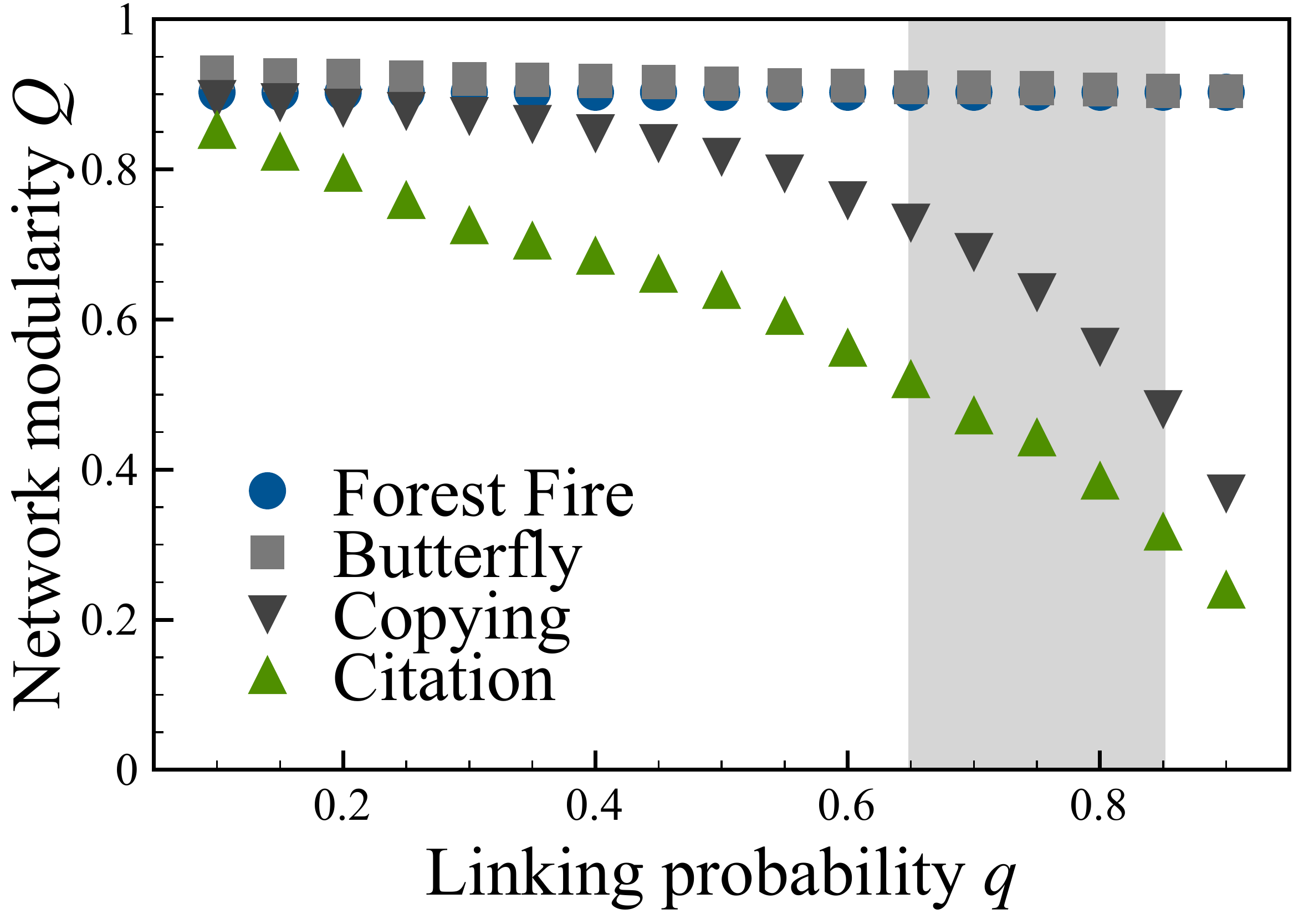}}
\caption{\label{fig:comparison}Comparison of graph models at different $\burning$ and $\linking=0.75$ (top), and $\burning=0.3$ and different $\linking$ (bottom). (Results are estimates of the mean over $100$ network realizations with $\n=1000$. See also caption of \figref{analysis}.)}
\end{figure*} 

Although \equsref{burned}{degree} are only valid in the limit of large network size, the bounds are rather tight for large enough $\n$ (see~\figref{analysis}). Thus, given network degree $\degree$ and fixed $\linking$, one can solve the system for $\burning$, which can be used for parameter estimation in practice (see~\secref{cora}).

\CIT~model generates small-world networks with scale-free degree distribution and community structure (see~\secref{analysis}). Furthermore, in contrast to \FF model, resulting networks are degree disassortative. We stress that the key factor here is that newly added nodes do not (necessarily) link to their ambassadors, which in fact produces degree assortativity. Since a node copies the links of its ambassadors, linking to them obviously promotes assortativity. However, in the absence of an explicit process introducing assortativity, (scale-free) networks are expected to be degree disassortative~\cite{JTMM10}. The analysis in~\secref{experiments} thus also includes a variant of \FF model denoted \BTF model, where a node links to its ambassadors only with probability $\linking$ (considered in~\cite{MAF08}), as well as a variant of the proposed \CIT model denoted \CPY model, where a node links to each ambassador~\cite{KR05} (for details see \figref{models}).

Other authors have proposed models very similar to ours~\cite{Vaz03,KR05,LKF07,MAF08,WH09}. Nevertheless, these either do not adopt the burning process to traverse the network or the model necessarily links the nodes to their ambassadors, which results in degree assortativity. More precisely, the set of the linked nodes is always a subset of the nodes burned (or vice versa). However, in the case of \CIT model, these two sets can intersect arbitrarily, while they can also be disjoint. 


\section{Experimental analysis} \label{sec:experiments}
\secref{analysis} conducts an empirical analysis of \CIT model and several alternatives proposed in the literature (see~\secref{model}). Next, networks constructed with different models are compared against a larger citation network (\secref{cora}).

\subsection{Analysis of the model} \label{sec:analysis}
\figref{comparison} shows basic statistics of the networks generated with different graph models for parameters $\burning$ and $\linking$ shown (see~\secref{model}). Most notably, only the proposed \CIT model gives degree disassortative networks measured by the mixing coefficient $\mixing\in[-1,1]$~\cite{New02} (see~\figsref{comparison:p:r}{comparison:q:r}). $\mixing$ is simply a Pearson correlation coefficient of degrees at links' ends. Thus, $\mixing\ll 0$ for \CIT model, while $\mixing\gg 0$ for \FF and \BTF models. Observe that \CPY model also generates networks with $\mixing<0$ for very large $\burning$ and $\linking$, however, these are much denser than comparable real-world networks (see~\figsref{comparison:p:k}{comparison:q:k}).

On the other hand, all models give small-world networks with short mean distance between the nodes $\distance$~\cite{AMO93} (see~\figsref{comparison:p:l}{comparison:q:l}) and high transitivity measured by the clustering coefficient $\clustering\in[0,1]$~\cite{WS98} (see~\figsref{comparison:p:C}{comparison:q:C}). Note that $\clustering$ increases with $\burning$, while $\linking$ has little effect on $\clustering$. Furthermore, all models generate networks with clear community structure according to modularity $\modularity\in[0,1]$~\cite{NG04}, where $\modularity$ is estimated using a fast multi-stage optimization~\cite{BGLL08} (see~\figsref{comparison:p:Q}{comparison:q:Q}). Although $\modularity$ decreases with increasing $\burning$ or $\linking$ in the case of \CIT and \CPY models, the values are somewhat comparable to those observed in real-world networks. \FF and \BTF models, however, appear to overestimate $\modularity$ for selected $\burning$ and $\linking$.

Networks constructed with \CIT model also reveal scale-free degree distributions~\cite{BA99} (see~\figref{cora:pk}), thus, the model generates most common properties of real-world networks. 

\begin{table}[b]
\centering
\caption{\label{tbl:cora}Comparison of \cora citation network and those constructed with different graph models for $\burning$ and $\linking$ shown. (Results are estimates of the mean over $100$ network realizations with $\n=23166$.)}
\begin{tabular}{cccccc} \HLINE
Model & $\burning$ & $\linking$ & $\m$ & $\degree$ & $\mixing$ \\\HLINE
\FF & $0.462$ & - & $88828$ & $7.669$ & $0.211$ \\\hline
\CIT & $0.369$ & $0.593$ & $89888$ & $7.760$ & $-0.047$ \\\HLINE
\cora & & & $\mathit{89157}$ & $\mathit{7.697}$ & $\mathit{-0.055}$ \\\HLINE
\end{tabular}
\end{table}

\subsection{Cora citation network} \label{sec:cora}
Due to a natural interpretation for citation networks (see \secref{model}), the proposed model has different practical applications in bibliometrics. We here analyze author citation dynamics based on the famous \cora dataset~\cite{MNRS00} that contains computer science papers collected from the web, and also the references automatically parsed from the bibliographies of the papers. We extract a citation network with $\n=23166$, while other statistics are reported in~\tblref{cora}.

\begin{figure}[b]
\centering
\subfigure[\label{fig:cora:pk}Distribution $P(\degree)$]{
\includegraphics[width=\plotwidth]{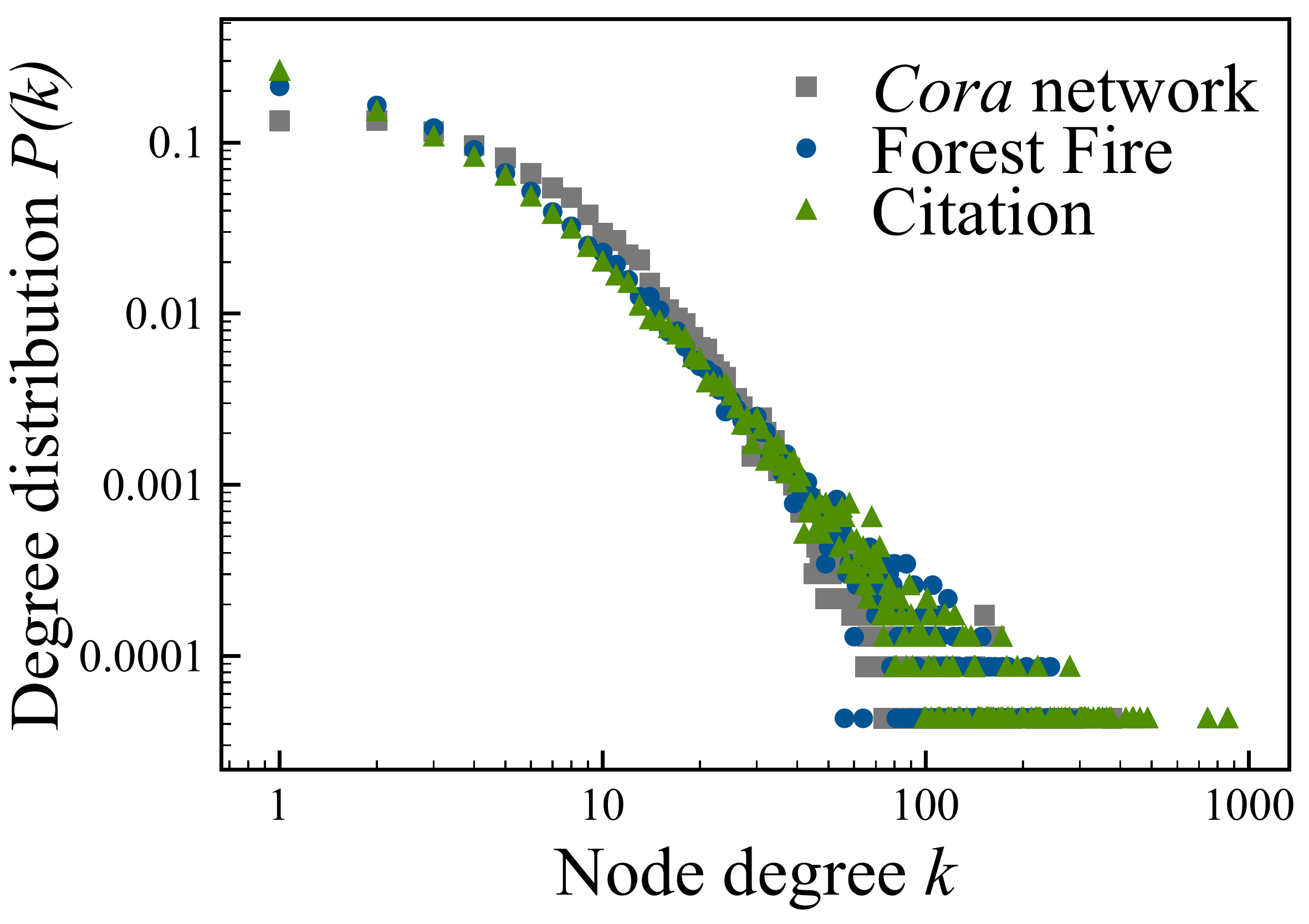}}
\subfigure[\label{fig:cora:kn}Neighbor degree $\ndegree$]{
\includegraphics[width=\plotwidth]{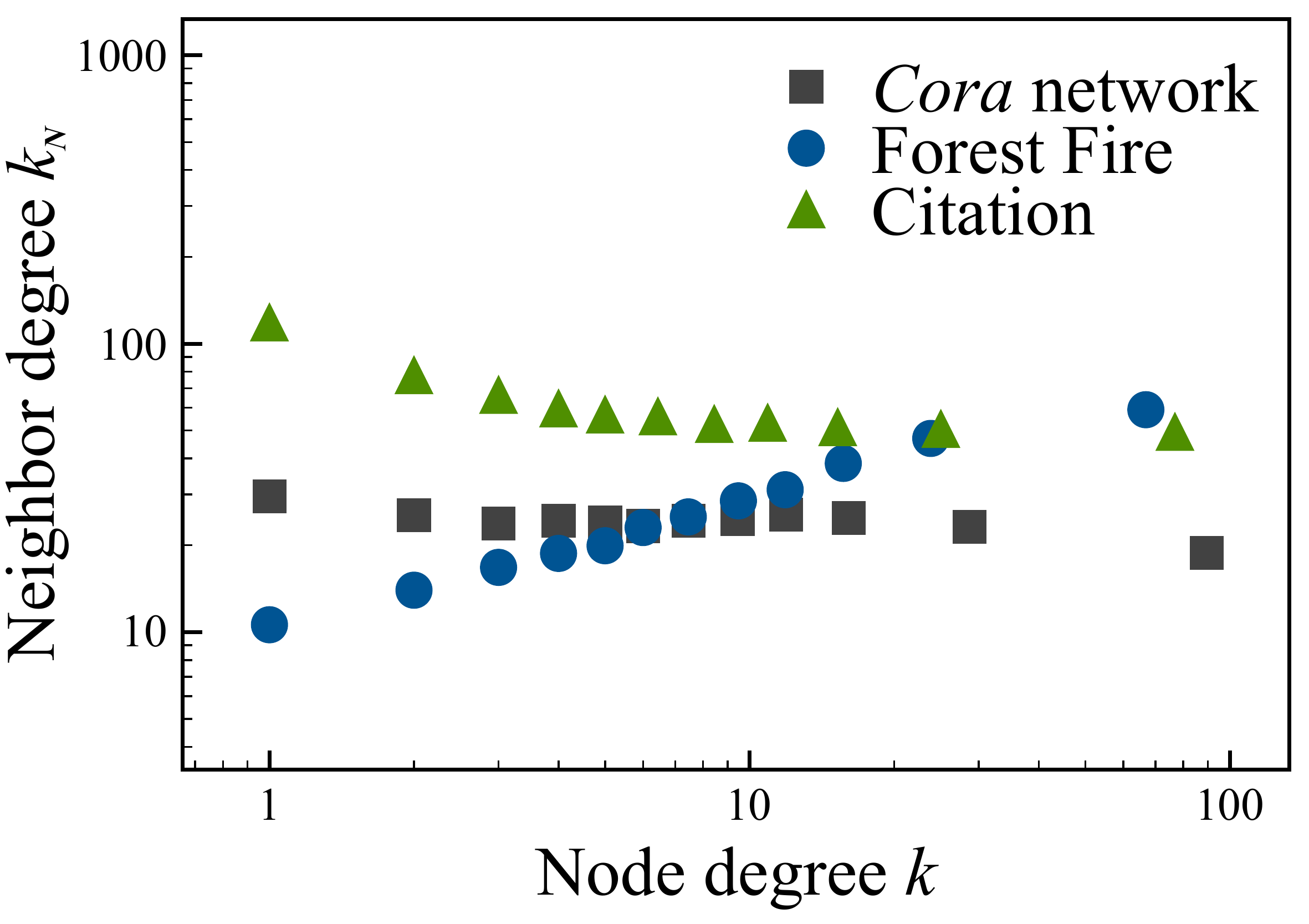}}
\caption{\label{fig:cora:k}Comparison of \cora network and those constructed with different graphs models (\tblref{cora}). Exponent $\alpha$ for a power-law fit $P(\degree)\sim k^{-\alpha}$ equals $3.3$ for \cora network and $2.5$ for graph models. (Nodes in~(b) are aggregated into equally-sized bins.)} 
\end{figure}

\tblref{cora} also includes the networks generated with \CIT and \FF models, where parameters $\burning$ and $\linking$ were estimated as described in~\secref{model}. Note that \CIT model well matches the disassortative mixing regime in \cora citation network (observe also a similar trend in~\figref{cora:kn}), while \FF model gives degree assortative networks. (For comparison based on other network properties see~\cite{SB12n}.)

Recall that $\burned$ in \equref{burned} can be seen as the number references actually read by an author of some paper. Thus, the fraction of papers considered by the authors, relative to the number of all papers cited, can be estimated to $2\burned/\degree=0.66$. The value is much larger than expected~\cite{SR03}, however, the results are largely influenced by an automatic sampling procedure~\cite{MNRS00} (i.e., on average, only $\degree/2=3.85$ references of each paper are also included in the network).


\section{Conclusion} \label{sec:conclusion}
The paper proposes a simple graph model that generates networks with most common properties of real-world networks and, in contrast to many other models, dissasortative degree mixing. The model also has a natural interpretation for citation networks with different practical applications.

Due to simplicity, the analysis in the paper is based on undirected networks. However, this presents a serious limitation, especially for citation networks considered here. Future work will extend the analysis to directed and also other types of networks, while more reliable datasets will be used for the analysis of author citation dynamics (based on \textit{DBLP} and \textit{WoS} data). Furthermore, the model will be rigorously compared against others with similar characteristics~\cite{TZJ09}.




\bibliographystyle{abbrv}

\begin{thebibliography}{10}

\bibitem{AMO93}
R.~K. Ahuja, T.~L. Magnanti, and J.~B. Orlin.
\newblock {\em Network flows: Theory, algorithms, and applications}.
\newblock Prentice-Hall, Upper Saddle River, {NJ}, 1993.

\bibitem{BA99}
A.~L. Barab{\'a}si and R.~Albert.
\newblock Emergence of scaling in random networks.
\newblock {\em Science}, 286(5439):509--512, 1999.

\bibitem{BGLL08}
V.~D. Blondel, J.-L. Guillaume, R.~Lambiotte, and E.~Lefebvre.
\newblock Fast unfolding of communities in large networks.
\newblock {\em J. Stat. Mech.}, P10008, 2008.

\bibitem{ER59}
P.~Erd{\H o}s and A.~R{\'e}nyi.
\newblock On random graphs i.
\newblock {\em Publ. Math. Debrecen}, 6:290--297, 1959.

\bibitem{FIA11}
G.~Facchetti, G.~Iacono, and C.~Altafini.
\newblock Computing global structural balance in large-scale signed social
  networks.
\newblock {\em P. Natl. Acad. Sci. {USA}}, 108(52):20953--20958, 2011.

\bibitem{GN02}
M.~Girvan and M.~E.~J. Newman.
\newblock Community structure in social and biological networks.
\newblock {\em P. Natl. Acad. Sci. {USA}}, 99(12):7821--7826, 2002.

\bibitem{HL11}
D.~Hao and C.~Li.
\newblock The dichotomy in degree correlation of biological networks.
\newblock {\em {PLoS} {ONE}}, 6(12):e28322, 2011.

\bibitem{JTMM10}
S.~Johnson, J.~J. Torres, J.~Marro, and M.~A. Mu{\~n}oz.
\newblock Entropic origin of disassortativity in complex networks.
\newblock {\em Phys. Rev. Lett.}, 104(10):108702, 2010.

\bibitem{KR05}
P.~L. Krapivsky and S.~Redner.
\newblock Network growth by copying.
\newblock {\em Phys. Rev. E}, 71(3):036118, 2005.

\bibitem{LJTBH11}
P.~J. Laurienti, K.~E. Joyce, Q.~K. Telesford, J.~H. Burdette, and S.~Hayasaka.
\newblock Universal fractal scaling of self-organized networks.
\newblock {\em Physica A}, 390(20):3608--3613, 2011.

\bibitem{LKF07}
J.~Leskovec, J.~Kleinberg, and C.~Faloutsos.
\newblock Graph evolution: Densification and shrinking diameters.
\newblock {\em {ACM} Trans. Knowl. Discov. Data}, 1(1):1--41, 2007.

\bibitem{MNRS00}
A.~K. {McCallum}, K.~Nigam, J.~Rennie, and K.~Seymore.
\newblock Automating the construction of internet portals with machine
  learning.
\newblock {\em Inf. Retr.}, 3(2):127{\textendash}163, 2000.

\bibitem{MAF08}
M.~{McGlohon}, L.~Akoglu, and C.~Faloutsos.
\newblock Weighted graphs and disconnected components: Patterns and a
  generator.
\newblock In {\em Proceedings of the {ACM} {SIGKDD} International Conference on
  Knowledge Discovery and Data Mining}, page 524{\textendash}532, New York,
  {NY}, {USA}, 2008.

\bibitem{New02}
M.~E.~J. Newman.
\newblock Assortative mixing in networks.
\newblock {\em Phys. Rev. Lett.}, 89(20):208701, 2002.

\bibitem{NG04}
M.~E.~J. Newman and M.~Girvan.
\newblock Finding and evaluating community structure in networks.
\newblock {\em Phys. Rev. E}, 69(2):026113, 2004.

\bibitem{NP03}
M.~E.~J. Newman and J.~Park.
\newblock Why social networks are different from other types of networks.
\newblock {\em Phys. Rev. E}, 68(3):036122, 2003.

\bibitem{SR03}
M.~V. Simkin and V.~P. Roychowdhury.
\newblock Read before you cite!
\newblock {\em Compl. Syst.}, 14:269--274, 2003.

\bibitem{SB11s}
L.~{\v S}ubelj and M.~Bajec.
\newblock Community structure of complex software systems: Analysis and
  applications.
\newblock {\em Physica A}, 390(16):2968--2975, 2011.

\bibitem{SB12n}
L.~{\v S}ubelj and M.~Bajec.
\newblock Clustering assortativity, communities and functional modules in
  real-world networks.
\newblock {\em e-print {arXiv:12082518v1}}, pages 1--21, 2012.

\bibitem{TZJ09}
L.~Tan, J.~Zhang, and L.~Jiang.
\newblock An evolving model of undirected networks based on microscopic
  biological interaction systems.
\newblock {\em J. Biol. Phys.}, 35(2):197--207, 2009.

\bibitem{Vaz03}
A.~V{\'a}zquez.
\newblock Growing network with local rules: Preferential attachment, clustering
  hierarchy, and degree correlations.
\newblock {\em Phys. Rev. E}, 67(5):056104, 2003.

\bibitem{VGB11}
S.~Vitali, J.~B. Glattfelder, and S.~Battiston.
\newblock The network of global corporate control.
\newblock {\em {PLoS} {ONE}}, 6(10):e25995, 2011.

\bibitem{WS98}
D.~J. Watts and S.~H. Strogatz.
\newblock Collective dynamics of 'small-world' networks.
\newblock {\em Nature}, 393(6684):440--442, 1998.

\bibitem{WH09}
Z.-X. Wu and P.~Holme.
\newblock Modeling scientific-citation patterns and other triangle-rich acyclic
  networks.
\newblock {\em Phys. Rev. E}, 80(3):037101, 2009.

\end{thebibliography}


\end{document}